\documentclass[twocolumn]{aastex631}
\usepackage{amsmath,amssymb}
\usepackage{graphicx}
\usepackage{tabularx}
\hypersetup{
    colorlinks=true,
    linkcolor=blue,
    citecolor=blue,
    filecolor=blue,
    urlcolor=blue
}

\usepackage{bm}
\usepackage{array}
\usepackage{cases}
\usepackage{amsmath}

\newcommand{\be}{\begin{equation}}
\newcommand{\ee}{\end{equation}}
\newcommand{\bea}{\begin{eqnarray}}
\newcommand{\eea}{\end{eqnarray}}
\newcommand{\dd}{\mathrm{d}}

\begin{document}

\shorttitle{Unmasking Stellar Feedback-Driven Bubbles}
\shortauthors{Angress et al.}

\title{Unmasking Stellar Feedback-Driven Bubbles: Identification and Properties Analysis}

\correspondingauthor{Aaron Angress}
\email{aaron.angress89@gc.cuny.edu}

\author[0009-0005-5250-9487]{Aaron Angress}
\affiliation{Department of Physics, Northeastern University, 100 Forsyth St., Boston, MA, 02115}
\affiliation{The Graduate Center of the City University of New York, 
365 5th Ave., New York, NY 10016}
\affiliation{Department of Physics and Astronomy, Lehman College of the CUNY, Bronx, NY 10468, US}
\affiliation{Center for Astrophysics $\mid$ Harvard \& Smithsonian, 
60 Garden St., Cambridge, MA, USA 02138}

\author[0000-0002-6747-2745]{Michael M. Foley}
\affiliation{Center for Astrophysics $\mid$ Harvard \& Smithsonian, 
60 Garden St., Cambridge, MA, USA 02138}

\author[0000-0002-4232-0200]{Sarah M. R. Jeffreson}
\affiliation{Center for Astrophysics $\mid$ Harvard \& Smithsonian, 
60 Garden St., Cambridge, MA, USA 02138}

\author[0000-0003-1312-0477]{Alyssa Goodman}
\affiliation{Center for Astrophysics $\mid$ Harvard \& Smithsonian, 
60 Garden St., Cambridge, MA, USA 02138}

\author{Lars Hernquist}
\affiliation{Center for Astrophysics $\mid$ Harvard \& Smithsonian, 
60 Garden St., Cambridge, MA, USA 02138}

\begin{abstract}
The identification and tracking of stellar feedback-driven galaxy bubbles is an important topic in star formation and galactic structure research. However, current observational analysis of bubbles is limited in scope; information on bubble lifetime is inaccessible. Simulation data thus provides a unique opportunity to glean some of these characteristics at high resolution. We present an investigation into the characteristics and evolution of hot, ionized bubbles in the interstellar medium of a dwarf spiral (NGC300-like) galaxy. We calculate the average radius, lifetime, temperature, density, and spatial distribution of the simulated feedback-driven bubbles using Lagrangian gas parcels, and we examine the relationship between these characteristics and the local galactic environment. We find exponential distributions of bubble lifetime and size, and we find a positive correlation between bubble lifetime and galactocentric radius. Finally, we predict how the data would appear in H$\alpha$ tracers and compare the simulated values to observations.  We find an additional positive correlation between the size of the bubbles and the galactocentric radius using their H$\alpha$ tracers.
\end{abstract}


\section{Introduction} \label{sec:intro}

Superbubbles are regions of hot, ionized gas in the interstellar medium (ISM) which may be produced by a variety of stellar feedback mechanisms, including stellar winds, radiation pressure, and supernovae \citep{WeaverBubbleModel, SilichRadPressure,TTSN}. Many galactic processes, from star formation to galactic outflows, are shown to depend on the frequency and morphology of superbubble formation \citep{Ceverino2009, Elmegreen2011, ZuckerLocal}. Extending from around 10 pc to 100s of pc in size \citep{BreitBubbleSizeExample} and lasting for 10s of Myr \citep{OchsendorfBubbleLifetime}, these cavities of hot gas can occupy a significant fraction of a galactic disk. Recent observations have shown that hundreds to thousands of superbubbles may concurrently exist in a single galaxy \citep{WatkinsMain}. 

 Superbubbles may have a number of distinct effects on the surrounding interstellar medium. They may impart large quantities of energy and momentum to the surrounding interstellar medium, dispersing and heating the gas, and thus preventing further star formation in a process known as negative feedback \citep{Elmegreen2011}. Conversely, superbubbles may condense gas into new star-forming regions at their surfaces (positive feedback). Positive feedback may localize SF in spiral arms and spurs \citep{ZuckerReview}. Understanding the creation, evolution, and impacts of superbubbles is crucial to characterizing galaxy formation and evolution. In this work, we distinguish between superbubbles, which refer to feedback-driven cavities containing hot, over-pressurized gas produced by clustered supernovae, and HI shells or supershells, which are colder, neutral structures that are often interpreted as the later evolutionary stages of superbubbles once the hot interior has cooled. HI shells and supershells have been extensively studied in the Milky Way over a wide range of galactocentric radii \citep{Heiles1979,Heiles1984,Hu1981,McClure-Griffiths2002}, demonstrating that such structures are not confined to the local ISM.

Due to new 3D dust mapping techniques and the incredible resolution offered by JWST, the spatial relationship between superbubbles and stellar associations can be probed like never before. In the Milky Way, nearly all of the local star formation appears to occur on the surfaces of expanding superbubbles \citep{ZuckerReview}. For example, the Local Bubble alone likely contributed to the formation of the Taurus, Corona Australis, Ophiuchus, Musca, Lupus, and Chamaelon star-forming regions \citep{ZuckerLocal}. The Perseus-Taurus Supershell contains the Perseus and Taurus star-forming regions, and the entire Orion star-forming complex appears to lie at the intersection of two superbubbles: the Orion Shell and the Orion-Eridanus superbubble \citep{Joubaud2019, Bialy2021, Foley2023}. Collectively, these superbubbles possess millions of solar masses of both dense molecular and atomic gas in their shells and hot ionized gas in their interiors, all within 500 pc of the sun.

Although superbubbles have been extensively studied within the galaxy, until recently, superbubbles have only been studied in a small number of extragalactic environments. Observations of HI stellar feedback shells in the Large Magellanic Cloud suggest that superbubbles likely drive production of new star-forming regions by increasing the molecular hydrogen fraction \citep{DawsonLMC}.  Similarly, 21 cm HI maps of nearby spiral and dwarf galaxies reveal large quantities of HI shells surrounding voids \citep{Bagetakos21cmMap}. With incoming JWST data (e.g. \cite{WatkinsMain}), extragalactic superbubble studies have entered a new era, allowing for unprecedented resolution in probing extragalactic substructure. These observations are time-critical, as early JWST results are beginning to reveal the detailed structure of superbubbles and their surroundings, offering fresh insight into whether stellar feedback can directly trigger new star formation.

However, despite these observational advances, a key limitation remains: observational studies alone cannot establish causality between superbubbles and triggered star formation.
Observational approaches can only statistically constrain the properties of superbubbles -- they cannot track the evolution of individual superbubbles over time. To study the effects of superbubbles on the interstellar medium and circumgalactic processes, simulations provide the unique opportunity to witness the direct effects of large samples of superbubbles as they grow and evolve. These data can reveal time-dependent properties of the superbubbles that are unavailable in observational data, like their lifetimes or radial profiles over time. Additionally, time-evolving data can prove whether superbubbles are creating new star formation regions, or if they are only sweeping up existing regions.

In this article, we present a study of a large population of time-evolving superbubbles in the galactic context with realistic ISM physics. We detail a procedure for identifying and tracking the time evolution of simulated superbubbles. Additionally, we analyze the lifetimes, size, spatial distribution, and observational potential for our identified superbubbles. Our sample of time-evolving bubbles will fill the knowledge gap concerning the lifetime and size evolution of these bubbles, especially with respect to their local galactic environment.

This paper is organized as follows: Section \ref{sec:sims} presents an overview of our simulations; Section \ref{sec:bubble identification} discusses our methodology for identifying and tracking bubbles; Section \ref{sec:bubble properties} presents properties of our superbubble sample; Section \ref{sec:Discussion} analyzes our model's caveats and its relationship with past and future work; and Section \ref{sec:Conclusions} summarizes our findings. 
\begin{figure*}
	\includegraphics[width=\linewidth]{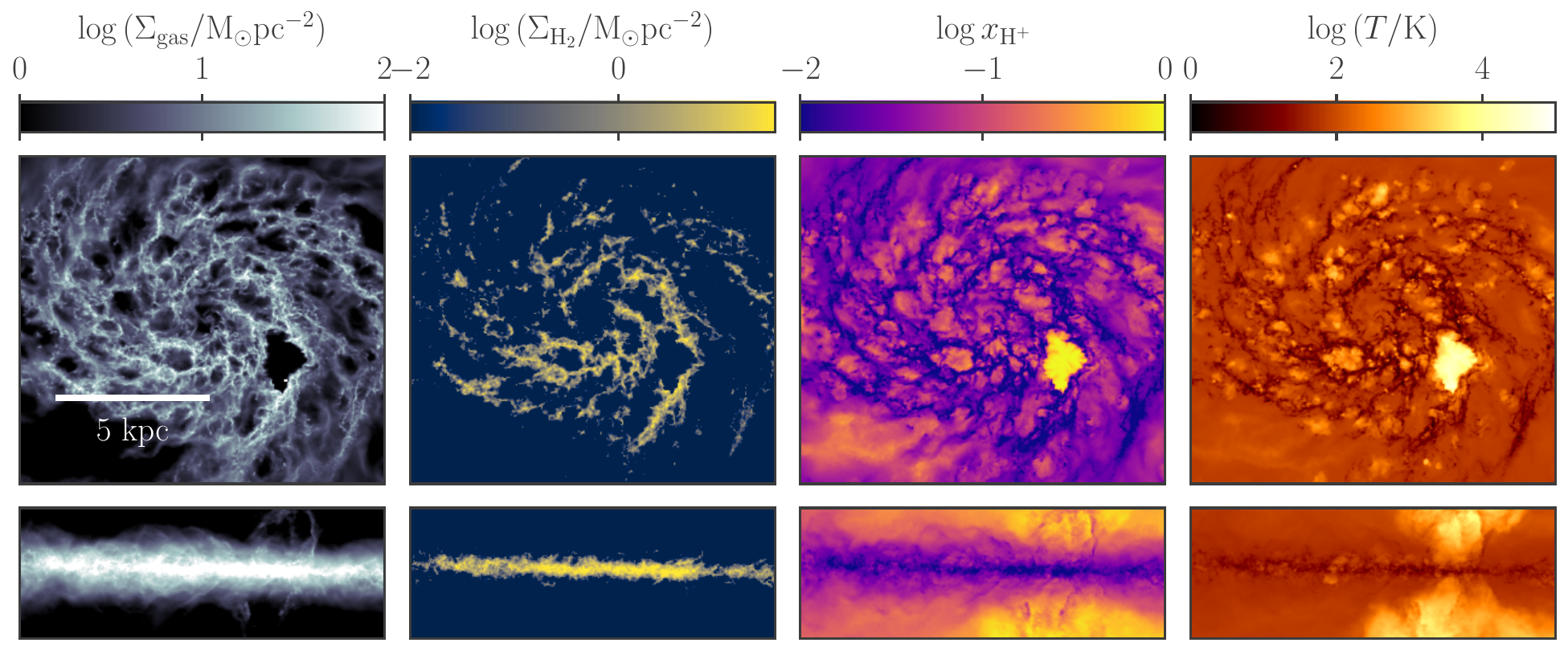}
	\caption{Column density maps of the total gas surface density ($\Sigma_{\rm gas}$, left), molecular gas surface density ($\Sigma_{\rm H_2}$, center-left), projected (averaged in the third direction) ionized hydrogen fraction ($x_{{\rm H}^+}$, center-right), and projected temperature ($T$, right) for the simulated dwarf spiral galaxy, viewed perpendicular to (top panels) and across (lower panels) the galactic mid-plane, at a simulation time of $800$~Myr.}
	\label{Fig::morphology}
\end{figure*}

\section{Simulations} \label{sec:sims}
We analyze the simulation of an NGC300 analog, first presented in~\cite{pubCloudsOfTheseus}. The spatial distribution of the total (left), ionized (center-left) and molecular (center-right) gas distributions is shown in Figure~\ref{Fig::morphology}, along with the projected gas temperature (right). As discussed in~\cite{pubCloudsOfTheseus}, the simulated galaxy reproduces the observable properties of NGC300 from molecular cloud to galaxy scales, including the molecular, atomic and stellar surface densities~\citep{2011MNRAS.410.2217W,Kruijssen2019}, the molecular and atomic depletion times~\citep{Kruijssen2019}, and the molecular gas surface densities and velocity dispersions of giant molecular clouds~\citep{Faesi18}. Here we provide a summary of the key features of our numerical method. We refer the reader to~\cite{pubCloudsOfTheseus} for a fuller explanation and for an overview of the physical properties of the simulated galaxy. 

\subsection{Hydrodynamics and N-Body Physics}
The simulation is run using the moving-mesh code {\sc Arepo}~\citep{Springel10}. The gaseous (hydrodynamical) component is represented by an unstructured moving mesh defined by the Voronoi tessellation around discrete points that move in accordance with local gas velocity, and a hybrid TreePM gravity solver is used to compute the gravitational acceleration vectors of the gas, stellar, and dark matter particles. Our NGC300-like initial condition has a mass resolution of $1.254 \times 10^7~{\rm M}_\odot$ for the dark matter particles, $3.437 \times 10^5~{\rm M}_\odot$ for the stellar particles, and $859~{\rm M}_\odot$ for the gas cells. The dark matter halo follows the profile of~\cite{Navarro97}, with a concentration parameter of $c=15.4$, a spin parameter of $\lambda=0.04$, a mass of $8.3\times 10^{10}~{\rm M}_\odot$ and a circular velocity of $V_{200} = 76~{\rm km~s}^{-1}$ at the virial radius. The stellar disc is of exponential form, with a mass of $1 \times 10^9~{\rm M}_\odot$, a scale-length of $1.39$~kpc, and a scale-height of $0.28$~kpc. The corresponding exponential gas disc extends beyond the stellar disc, with a mass of $2.2 \times 10^9~{\rm M_\odot}$ (giving a gas fraction of 68~per~cent) and a scale-length of $3.44~{\rm kpc}$. The molecular scale-height is $50~{\rm pc}$.

Throughout the runtime of the simulations, the adaptive gravitational softening scheme in {\sc Arepo} is employed with a softening length set to $1.5$ times the Voronoi gas cell size, reaching a minimum of $3$~pc. The softening length is fixed at $3$~pc for stellar particles and at~$260$~pc for dark matter particles, based on the convergence tests of~\cite{2003MNRAS.338...14P}. Given that the gas disk scale height and Toomre mass are resolved at all scales in our simulations, the adaptive softening scheme ensures that minimal artificial fragmentation occurs at scales larger than the Jeans length~\citep{Nelson06}.

\subsection{Thermodynamics}
We set the initial gas temperature in our simulation to $10^4~{\rm K}$. This re-equilibrates on a time-scale of $\la 10~{\rm Myr}$ to a state of thermal balance between line-emission cooling and heating due to photo-electric emission from dust grains and polycyclic aromatic hydrocarbons, modeled using the simplified network of hydrogen, carbon, and oxygen chemistry introduced in~\cite{NelsonLanger97,GloverMacLow07a,GloverMacLow07b}. The network computes and tracks fractional abundances for the species ${\rm H}$, ${\rm H}_2$, ${\rm H}^+$, ${\rm He}$, ${\rm C}^+$, ${\rm CO}$, ${\rm O}$ and ${\rm e}^-$ for each gas cell. This chemistry is self-consistently coupled to the heating and cooling of the interstellar medium, according to the atomic and molecular cooling function of~\cite{Glover10} (see their Table 1 for a full list of included heating and cooling processes). The thermal evolution of the gas in our simulations therefore depends on the gas density and temperature, as well as on the strength of the interstellar radiation field, the cosmic-ray ionization rate, the dust fraction and temperature, and the set of chemical abundances tracked for each gas cell. We take a value of $1.7$~Habing fields for the UV component of the ISRF~\citep{Mathis83} and a value of $3 \times 10^{-17}$~s$^{-1}$ to the cosmic ionization rate~\citep{2000A&A...358L..79V}. We assume the solar neighborhood value of the dust-to-gas ratio. Finally, we model the dust- and self-shielding of molecular hydrogen from dissociation by the ISRF using the {\sc TreeCol} algorithm introduced by~\cite{Clark12}.

\subsection{Star Formation}
The star formation rate volume density in our simulation is given by
\begin{numcases} {\frac{\dd \rho_{*,i}}{\dd t} = }  \label{Eqn::SF1}
\frac{\epsilon_{\rm ff} \rho_i}{t_{{\rm ff},i}}, \; \rho_i \geq \rho_{\rm thresh}, T_i \leq T_{\rm thresh} \\
0, \; \rho_i < \rho_{\rm thresh}, T_i > T_{\rm thresh}
\end{numcases}
where $t_{{\rm ff}, i} = \sqrt{3\pi/(32 G\rho_i)}$ is the local free-fall time-scale for the gas cell $i$ with a mass volume density of $\rho_i$, and $\epsilon_{\rm ff}$ follows the parametrization of~\cite{Padoan17}, such that
\begin{equation} \label{Eqn::SF2}
\epsilon_{\rm ff} = 0.4 \exp{(-1.6 \alpha_{\rm vir}^{0.5})}.
\end{equation}
The virial parameter $\alpha_{\rm vir}$ on cloud scales is computed during simulation run-time within over-dense regions surrounding each star-forming gas cell, determined via a variation of the~\cite{1960mes..book.....S} approximation used in~\cite{2020MNRAS.495..199G}. We set an upper limit of $T_{\rm thresh} = 100{\rm K}$ on the temperature below which star formation is allowed to occur, and a lower limit of $\rho_{\rm thresh}/m_{\rm H} \mu = 100~{\rm cm}^{-3}$ on the density, where $\mu$ is the mean mass per H atom.

\subsection{Supernova Feedback}
Each resulting star particle is assigned a stellar population drawn stochastically from a~\cite{Chabrier03} initial stellar mass function (IMF) via the Stochastically Lighting Up Galaxies (SLUG) stellar population synthesis model~\citep{daSilva12,daSilva14,Krumholz15}. By evolving each stellar population along Padova solar metallicity tracks~\citep{Fagotto94a,Fagotto94b,VazquezLeitherer05} while using {\sc Starburst99}-like spectral synthesis~\citep{Leitherer99}, SLUG provides an ionizing luminosity for each star particle at each simulation time-step, as well as the number of supernovae $N_{*, {\rm SN}}$ it has generated and the mass $\Delta m_*$ it has ejected.

Using the values of $N_{*, {\rm SN}}$ and $\Delta m_*$ provided for each star particle by our stellar evolution model, we model the momentum and thermal energy injected by supernova explosions at each simulation time-step. If $N_{*, {\rm SN}} = 0$, then we assume that any mass loss results from stellar winds. If $N_{*, {\rm SN}} > 0$, we assume that all mass loss results from supernovae. Our simulations do not resolve the energy-conserving/momentum-generating phase of supernova blast-wave expansion, so we explicitly inject the terminal momentum of the blast-wave to avoid over-cooling, as discussed in~\cite{KimmCen14}. We use the unclustered parametrization of the terminal momentum injected into the gas cells $k$ that share faces with a central cell $j$, derived from the high-resolution simulations of~\cite{Gentry17}, and given by
\begin{equation} \label{Eqn::Gentry17}
\frac{p_{{\rm t}, k}}{{\rm M}_\odot {\rm kms}^{-1}} = 4.249 \times 10^5 N_{j, {\rm SN}} \Big(\frac{n_k}{{\rm cm}^{-3}}\Big)^{-0.06},
\end{equation}
where $N_{j, {\rm SN}}$ is the total number of supernovae associated with all star particles for which gas cell $j$ is the nearest neighbor. We distribute this terminal momentum to the gas cells surrounding the central cell, as described in~\cite{2020arXiv200403608K,2021MNRAS.505.3470J}. The upper limit on the terminal momentum is set by kinetic energy conservation as the shell sweeps through the gas cells $k$~\citep[see also][for similar prescriptions]{Hopkins18,Smith2018}.

\subsection{HII Regions}
In addition to supernova feedback, we include pre-supernova feedback from HII regions, following~\cite{2021MNRAS.505.3470J}. This model takes account of the momentum injected by both radiation pressure and the thermal pressure from heated gas inside the HII region, according to the analytic work of~\cite{Matzner02,KrumholzMatzner09}. We group the star particles in the simulation via a Friends-of-Friends prescription of linking length equal to the HII region ionization front radius, improving the numerical convergence of the feedback model. The momentum is distributed to the gas cells that adjoin a central cell closest to the center of the luminosity of each Friends-of-Friends group. The gas cells inside the resulting grouped Str\"{o}mgren radii are also heated self-consistently and are not allowed to cool below a temperature floor of $7000$~K, for as long as they receive ionizing photons from the star particles. We rely on the chemical network to ionize the gas in accordance with the thermal energy injected, and so do not explicitly adjust the chemical state of the heated gas cells. \cite{2021MNRAS.505.3470J} finds that the ratio of energy pumped into the ISM by supernovae compared to that of presupernova-feedback is about 100:1. Similarly, the ratio of momentum injected by supernovae compared to presupernova-feedback is about 20:1.

Finally, we include passive tracer particles in the simulation, which allows us to track the Lagrangian mass flow of gas, following the Monte Carlo prescription of~\cite{2013MNRAS.435.1426G}. The effective tracer particle mass between the simulation times of $500$ and $800$~Myr analyzed in this work is stable at a value of $\sim 450~{\rm M_\odot}$, corresponding to $1.9$ tracer particles per gas cell. Tracer particles are moved along with the gas cells in the simulation and are exchanged between cells according to a probability set by the mass flux between them. When a gas cell is converted to a star particle, the tracer particles associated with that gas cell are moved to the star particle with a probability set by the ratio of the stellar mass to the original gas cell mass. Similarly, tracer particles attached to star particles are transferred back to the gas reservoir according to the masses ejected in stellar winds and supernovae. As such, the mass of tracers in the gas and stellar reservoirs remains equal to the masses of these reservoirs throughout the simulation.

\section{Bubble Identification} \label{sec:bubble identification}
\begin{figure*}
    \centering
    \includegraphics[width=\textwidth]{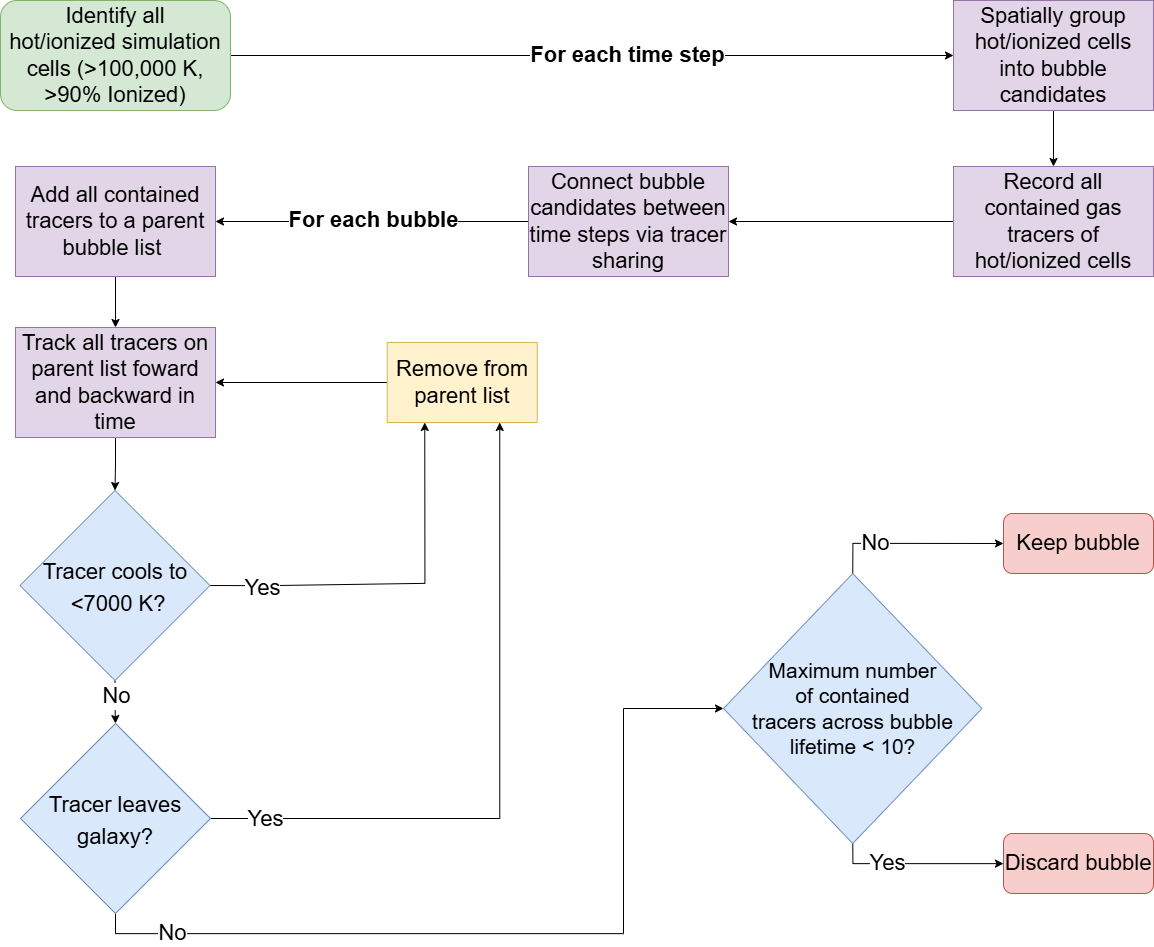}
    \caption{Bubble identification and tracking flowchart.}
    \label{Fig::flowchart}
\end{figure*}

In this section, we describe the bubble identification and time-evolution tracking procedure. In summary, the bubble identification algorithm groups hot, ionized gas cells into spatially connected groups, links gas tracers to them, and tracks the tracers as they heat and cool due to supernova feedback. See the flowchart in Figure~\ref{Fig::flowchart} for an overview of the bubble identification and tracking process.

\subsection{Physical Identification}\label{sec:physical identification}

We begin identification of bubble candidates by finding all the hot and ionized cells indicative of supernova feedback. At each time step, we identify all gas cells within the galaxy $(R \le 7$~kpc$, |z| \le 0.350$~kpc) with a temperature $T_{cell} > 100,000$~K and ionization fraction $X_{HP} > 0.9$. These temperatures and ionization fractions are only attainable in this simulation via supernova feedback. We use Cosmic Lyman-alpha Transfer (COLT)---a Monte Carlo radiative transfer solver for post-processing hydrodynamical simulations \citep{COLT}---to reconstruct the Voronoi mesh for the simulation at intervals of 1~Myr, and so to identify the gas cells that are neighbors within the mesh (those which share a cell face). Using this information, we recursively group the hot, ionized cells together into distinct, spatially connected bubble candidate groups. We do not limit the size of the candidate bubbles. We find an average of 435 of these candidates per time-step. We record the gas tracers contained in the cells of each of these candidates for use later, and we repeat this process for all simulation times in our analysis window, $t = [500, 800]$~Myr.

\subsection{Time Evolution Tracking}\label{sec:time evolution tracking}

Following the identification of individual superbubble candidates (groups of spatially connected, hot, highly ionized simulation cells at a single time step), we link these bubble candidates across time steps to identify time-evolving bubbles for further analysis. We store these connections between candidates using NetworkX graphs. We begin with two adjacent time steps, $t_1$ and $t_2$. If at least one gas tracer is shared between a bubble candidate in $t_1$ and a candidate in $t_2$, we link the candidates together by recording a connection between them. We repeat this linking procedure for all 301 time steps and create a network of connections between candidates across consecutive time steps. If two candidates in $t_1$ share tracers with one candidate in $t_2$, we label this as a bubble merger event. If one candidate in $t_1$ shares gas tracers with two candidates in $t_2$, we label this as a split event. While these events are important for individual bubble evolution, they are not essential for the statistical analysis we present in the rest of this work. We record no connections between bubble candidates in nonadjacent time steps.

Following this procedure, we record thousands of distinct, non-interacting sections of the time-evolution network that we label as individual time-evolving bubbles. No two distinct bubbles may share the same tracers at a single time step. For each bubble, we iterate across all its constituent cells. We add the contained gas tracers of every cell of the bubble to a tracking list unique to the current bubble. At each time step, we remove from the list all tracers that have traveled outside the radius of the galaxy ($R > 7$~kpc), left the original galactic z-cut ($|z| > 0.350$~kpc), or dropped below 7000~K. We set these spatial thresholds to remove stray tracers that remain hot and ionized in non-physical regions outside the simulated galaxy. 

After we have processed all time steps connected by tracer-sharing within a bubble, we continue to filter out gas particles from the tracking list by the same thresholds in the forward and reverse time directions, in effect tracking the heating and cooling of the gas outside of the hot, connected candidate phase. We additionally remove any tracers that are contained in the hot candidates of other bubbles to only track gas that has been influenced by the bubble in question. Without this filter, some gas may be reheated above 7000~K multiple times by other bubbles, artificially extending the lifetime and influencing the analysis of properties of the tracer's original bubble. We continue removing tracers from the list until there are less than 2 tracers remaining, or we have reached the end of available simulation data. When there are no longer at least two tracers, we cannot calculate the size of the bubble.

For every time-evolving list of tracers of each bubble, we calculate the maximum instantaneous number of tracers in the tracking list. We consider bubbles with a maximum of $<$ 10 tracers to be unresolved and remove them from the analysis. This resolution limit imposes a radial resolution on the bubbles we can identify (70~pc) which we will discuss in Section \ref{subsubsec:size resolution}. This threshold is partially arbitrary; we find that this limit strikes a balance between capturing bubbles with enough tracers to average across their properties while also identifying bubbles with smaller sizes. This size resolution also demands supernovae for feedback sources of the bubbles, as other sources of feedback alone could not produce similarly sized bubbles. We use the remaining bubbles for the final analysis. These bubbles may contain multiple hot, ionized candidate substructures (each composed of their own groups of cells) that merge and split at a single time step. We conduct property analysis of the bubbles as an average across all of the contained tracers in a bubble.

We emphasize that our bubble identification algorithm is intentionally designed to operate utilizing gas temperatures and without reference to the locations or timing of individual supernovae. As such, it identifies physical ISM structures rather than individual feedback events. While this implies that non-supernova-powered structures are not detected by construction, it also allows for closer comparison to observational bubble catalogs, which similarly lack information about the underlying supernova distribution. In this framework, a single detected bubble may correspond to multiple supernovae, and some supernovae may not be associated with a distinct bubble. We consider supernovae rates in relation to bubble formation rates in Section~\ref{subsec:bubble formation rate}.

Finally, we designate a subset of tracers for each bubble as H$\alpha$-emitting tracers. We define H$\alpha$-emitting tracers as any tracer that has a temperature $T_{tracer} =~[7000, 10000]$~K. These temperatures are typically associated with H$\alpha$ emission in ISM observations \citep{DraineISM}. Because H$\alpha$ is found along the edge of bubbles in observations, it is important to define this subset of our data for comparison \citep{WatkinsMain}.

\section{Bubble Properties} \label{sec:bubble properties}

In this section, we present the time-evolving properties of the 2778 identified bubbles, including lifetime, size, temperature, density, and galactocentric radius. We additionally search for relationships between these properties and provide synthetic H$\alpha$-observables for the bubbles.

\subsection{Bubble Formation Rate}
\label{subsec:bubble formation rate}

\begin{figure}[ht!]
	\includegraphics[width=\linewidth]{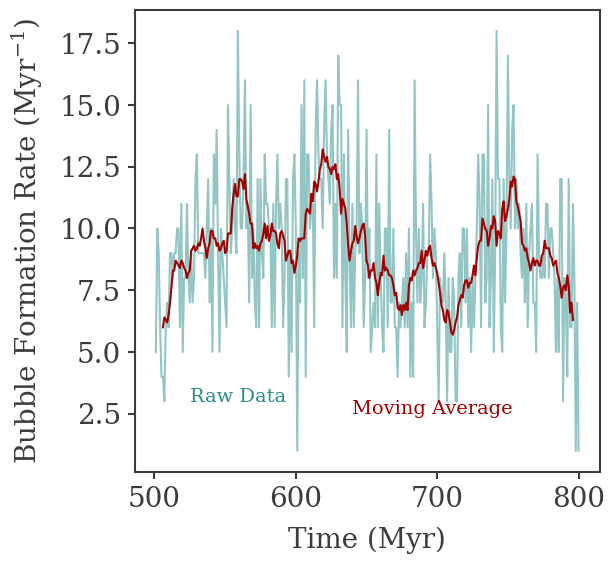}
	\caption{Bubble formation rate vs. time. The raw data appears in teal and the moving average (window size = $10$~Myr) appears in red for visual clarity.}
	\label{Fig::bubble formation rate}
\end{figure}

We calculate the bubble formation rate by counting the number of new bubbles that appear at each time step. We plot this value vs. simulation time in Figure~\ref{Fig::bubble formation rate}. The average bubble formation rate across all the simulation data is about 9/Myr.

Meanwhile, the global galactic star formation rate for our galaxy is on the order of $0.1-0.5$~$\frac{M_{\odot}}{yr}$, similar to observed SFRs for NGC 300 \citep{HelouNGC300,MondalNGC300,BinderNGC300}. More than 900,000 supernovae occurred throughout the total available simulation time, averaging out to around 3000 supernovae per Myr. Given the global SFR, this supernova rate appears reasonable, as we expect about 1 supernova for every 100$M_{\odot}.$ However, because we are only identifying around 9 new bubbles per Myr, this suggests that we are only detecting the largest bubbles. Many supernovae likely contribute to bubbles that are smaller than our resolution limit, which we discuss in Section~\ref{subsubsec:size resolution}. Similarly, to create the larger superbubbles we identify, clustered supernovae are likely necessary, leading to multiple supernovae counting towards a single bubble, lowering the ratio of bubbles to supernovae.

\subsection{Lifetime} \label{subsec:lifetime}

\begin{figure}[ht!]
	\includegraphics[width=\linewidth]{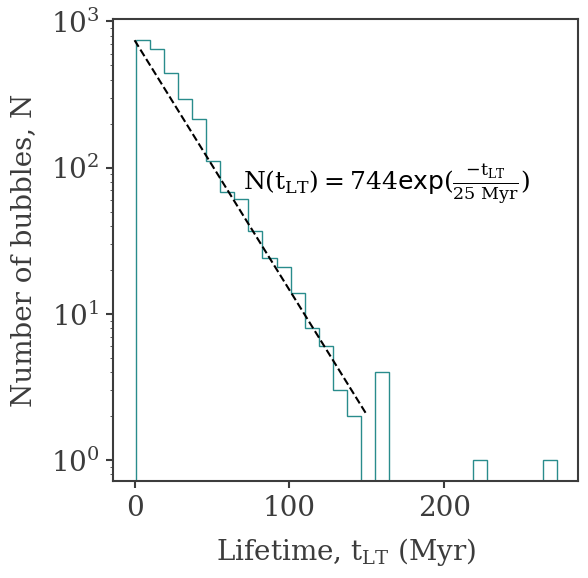}
	\caption{Bubble lifetime distribution. The purple dashed line indicates the exponential decay of the distribution as a function of lifetime in Myr ($t_{LT}$). Its slope is -25~Myr$^{-1}$. The magnitude of the slope is the mean lifetime of the bubbles within the domain of the fitted curve, with a subsequent half-life of 18~Myr. Long-lived bubbles toward the right end of the histogram may be highly-connected superbubbles.}
	\label{Fig::lifetime}
\end{figure}

We calculate the lifetime of the bubbles as the time difference between the appearance of the first identified tracers heating up before the hot phase of the bubbles' lifetimes and the disappearance of the last identified tracers cooling down after the hot phase. We produce the lifetime distribution in Figure~\ref{Fig::lifetime}. We remove the bubbles that exist at the first and last available simulation times ($t = 500$ and $800$~Myr, respectively; $\approx2\%$ of all bubbles) because information about their lifetime is clearly limited---some may start before or die after the available simulation endpoints, so their true lifetime is hidden. The remaining bubbles live their full lifetimes within available simulation data.

We fit an exponential function to the data, ${\rm Count}(t_{LT}) = 744\exp(\frac{-t_{LT}}{25})$ with $t_{LT}$ (lifetime) measured in Myr and find a half-life of 18~Myr. The slope of the trend line is $-25~{\rm Myr}^{-1}$, suggesting a mean lifetime of 25~Myr. 

There are 6 bubbles that fall outside the expected exponential distribution of lifetimes. Upon further investigation, we find that these bubbles are amongst those containing the highest maximum number of tracers during their lifetime, in the hundreds or thousands compared to the vast majority of bubbles that contain 10 to 50 tracers. Because of their high tracer content, these superbubbles are more likely to extend their lifetimes through mass sharing with other hot candidates. These bubbles also have sizes on kpc scales. We will explore this further in the following discussion of bubble size.

\subsection{Size} \label{subsec:size}
\begin{figure}[ht!]
	\includegraphics[width=\linewidth]{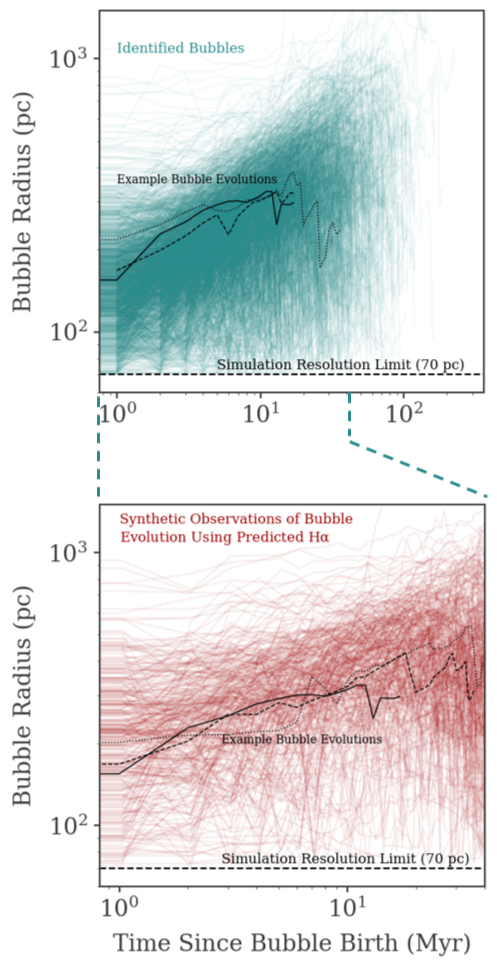}
	\caption{Bubble size evolution (top) and predicted observable size evolution using synthetic  H$\alpha$ (bottom). The blue indicates the bubble size evolution while the pink indicates the size evolution of only the gas predicted to be involved in H$\alpha$ emission. The synthetic observations take place only up to 40~Myr after bubble birth as prescribed by \cite{WatkinsMain}. The bottom dashed black line denotes our 70~pc resolution limit. The other black lines are example bubble size evolutions.}
	\label{Fig::size}
\end{figure}

We define the center of a bubble at a given time step as the centroid of its contained tracers. The average radius of the bubble is taken as the average of the distances of each of the tracers to the center of the bubble. We calculate this radius for each time step of the lifetime of each bubble and plot their size evolution in Figure~\ref{Fig::size}. The bottom black dashed line denotes the size resolution for the smallest bubbles we can identify. The mass resolution of the simulation imposes a size resolution that limits the range of bubbles that we can calculate a size for (see section~\ref{subsubsec:size resolution}). We find a radial resolution limit of 70~pc, which is denoted in Figure~\ref{Fig::size} by the bottom dashed line on each plot.

\begin{figure}[ht!]
	\includegraphics[width=\linewidth]{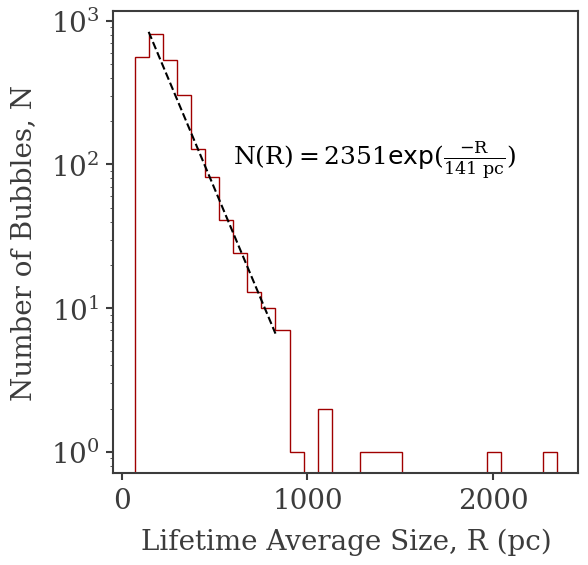}
	\caption{Bubble size distribution. We calculate the average size of each bubble across its first 40~Myr of lifetime using the H$\alpha$ tracers. The dashed line indicates the exponential decay of the distribution as a function of bubble radius (R). Its slope is -141~pc$^{-1}$. The population mean is 244~pc.}
	\label{Fig::size dist}
\end{figure}

We additionally create an analog for the predicted observable H$\alpha$ emission from the simulated gas by selecting tracers with temperature $T_{tracer} = [7000, 10000]$~K. We recalculate the size evolution of these bubbles using only the H$\alpha$ up to 40~Myr after their birth (if they survive at least 40~Myr) as prescribed by \cite{WatkinsMain}. These are plotted in magenta in the bottom plot in Figure~\ref{Fig::size}. We also average the size of each bubble as calculated using its H$\alpha$ tracers over its lifetime to create the bubble size distribution in Figure~\ref{Fig::size dist}

As expected, we find that the bubbles expand in size as they age. The size distribution of the average size of the bubbles over their lifetime is predominantly exponential. Supernovae drive the expansion of the bubbles' gas. The lower size limit of the majority of the identified bubbles nears the resolution limit we set.

A small number of extremely large superbubbles grow to kpc scales. We propose that these outliers simply emerge from our tracking procedure for the bubbles. We initially connect the hot bubble candidates without limiting the total size of the connected structures. If we identify a very large structure, then it is likely to contain large amounts of tracers. This excess number of tracers connects the structure to even more hot candidates in subsequent time steps, leading to superbubbles that span kpc scales in the galaxy. As we add more tracers to the parent list of the bubble, some may be dragged in opposite directions from each other, increasing the size of a bubble. In an observation, such a large structure may appear as many touching and nested bubbles.

\subsection{Temperature and Density} \label{subsec:temperature and density}
\begin{figure}[ht!]
	\includegraphics[width=\linewidth]{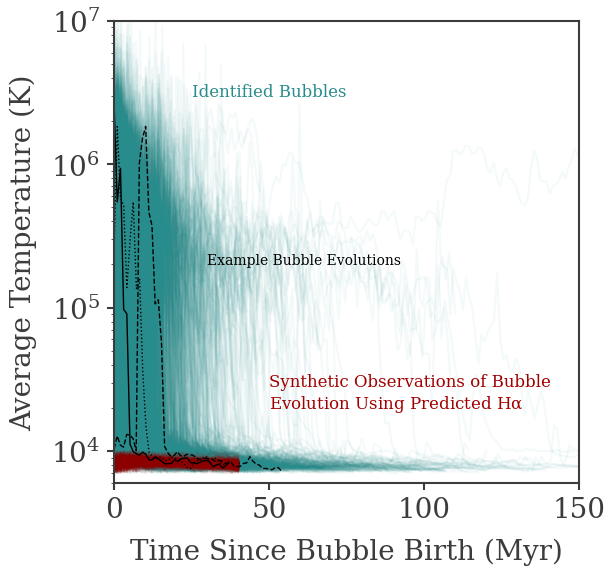}
	\caption{Bubble temperature evolution. Most bubbles experience a rapid drop in temperature before 20 Myr, while some are reheated by additional supernovae. H$\alpha$ emitting gas appears exactly as defined in Section~\ref{subsec:size}. The black lines are example bubble temperature evolutions.}
	\label{Fig::temperature}
\end{figure}

We calculate the average temperature of the bubbles as the average temperature of the contained tracers and plot the temperature evolution of each bubble in Figure~\ref{Fig::temperature}. We follow a similar procedure to represent the gas involved in H$\alpha$ emission as in Section~\ref{subsec:size}. Note the rapid drop in the average temperature of the majority of the bubbles by 20~Myr after their birth, corresponding to the end of sustained supernova heating. Because the bubbles begin to die as their gas cools below $7000$~K, this drop likely influences the half-life measured in the lifetime of the bubbles. However, some bubbles remain hot or begin heating again after 20~Myr. It is possible that these bubbles have new supernovae that drive further heating and expansion later into their lifetime. 

\begin{figure}[ht!]
	\includegraphics[width=\linewidth]{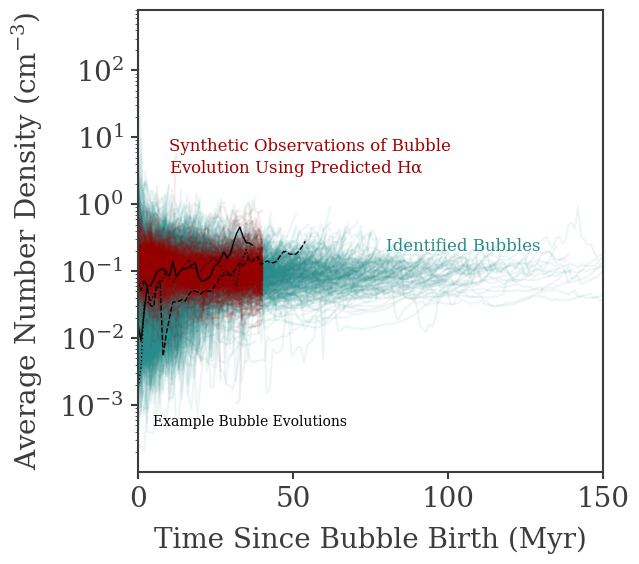}
	\caption{Bubble particle number density evolution. Most bubble densities evolve very gradually across their lifetime, staying within the range of $10^{-2}$ to $10^0~cm^{-3}$. H$\alpha$ tracer densities trend toward the upper end of this range, as expected of traditionally observed H$\alpha$ rings around bubble voids. The black lines are example bubble temperature evolutions.}
	\label{Fig::density}
\end{figure}

We calculate the average particle number density of the bubbles as the average number density of the contained tracers and plot the density evolution of each bubble in Figure~\ref{Fig::density}. We find that the average particle density of the majority of the bubbles falls in the range of $10^{-2}$ to $10^0~cm^{-3}$. Given the data in \ref{Fig::density}, we use $0.092~cm^{-3}$ as an approximation for the density of the interior of a bubble in Equation~\ref{Eqn::Resolution}. Overall, bubble particle number density appears relatively constant across the bubbles' lifetimes as compared to their other properties. 

\subsection{Porosity}

\label{subsec:porosity}
\begin{figure}[ht!]
	\includegraphics[width=\linewidth]{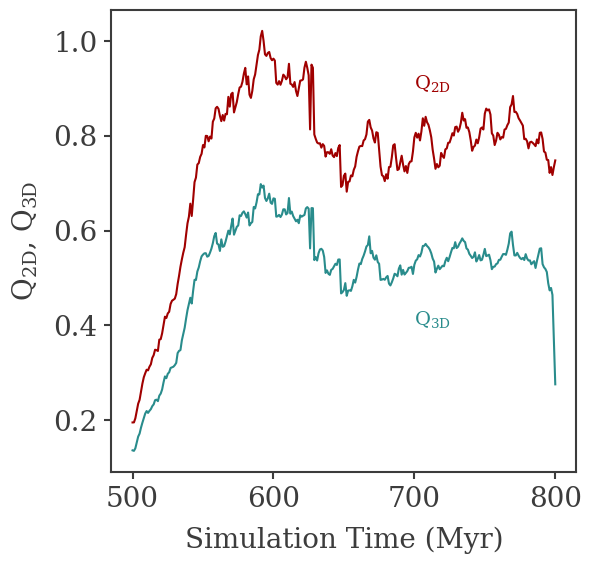}
	\caption{$Q_{2D}$ and $Q_{3D}$ vs. time. 3D bubble volumes are approximated to spheres with radii calculated in Section~\ref{subsec:size}}. 2D Bubble areas are approximated to circles with radii calculated using face-on radius (x-y plane radius).
	\label{Fig::porosity}
\end{figure}

Using the size of the bubbles, we can calculate the porosity parameters $Q_{2D}$ and $Q_{3D}$ of the galaxy as described in \cite{Porosity}. For every time step, we find the fractional area and volume of the galaxy occupied by bubbles approximated as circles and spheres, respectively. We plot these parameters with respect to simulation time in Figure~\ref{Fig::porosity}.

These are rather high porosity parameter values, especially for $Q_{2D}$. $Q_{2D}$ even briefly exceeds 1 near $t=600$~Myr. This indicates a high level of overlap between the bubbles, as overlapping areas and volumes were not removed for this analysis. For instance, two bubbles may experience their primary hot phases in the same set of time steps, but they are spatially disconnected. We track each bubble's contained tracers as they cool following this hot phase. In future time steps, the tracers of each bubble may expand past tracers of other bubbles as they cool, creating overlap. These results are likely dominated by the largest identified superbubbles. As seen in in Section~\ref{subsec:size}, the largest bubbles grow to kpc size scales, comparable to the entire size of the galaxy, so these results are expected.

\subsection{Energy}
\label{subsec:energy}

\begin{figure}[ht!]
	\includegraphics[width=\linewidth]{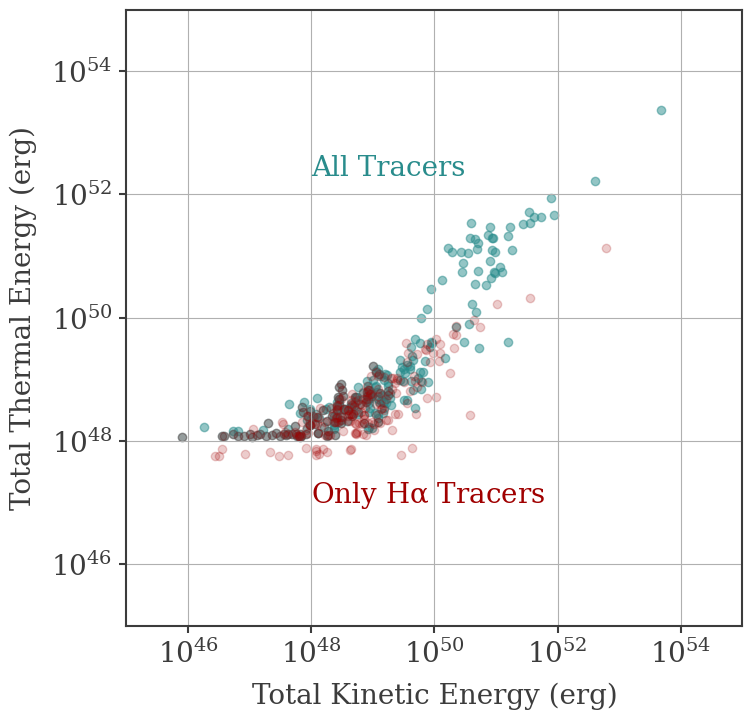}
	\caption{Kinetic and thermal energy for all identified bubbles at $t = 700$~Myr using all tracers (teal) and only H$\alpha$ tracers (red).}
	\label{Fig::energy}
\end{figure}

At $t=700$~Myr, we calculate the total thermal energy of each bubble, $\Sigma_i  m_{t}u_i$, where $m_{t}$ is the mass of the tracers ($450$~$M_{\odot}$) and $u_i$ is the specific internal energy of tracer i contained within the bubble. We proceed similarly to calculate the kinetic energy of each bubble with respect to its center of mass. We repeat the same process but only using the H$\alpha$ tracers, and we plot both pairs of energies against each other in Figure~\ref{Fig::energy}.

Both the energies of only the H$\alpha$ tracers and all the tracers follow similar trends with one obvious difference--when we utilize all of the tracers, not just H$\alpha$, the range of total thermal energy is much wider. The tracers must be above the minimum temperature threshold to be classified as a bubble, and the temperature is related to the specific internal energy, so the number of contained tracers has a large influence over the total thermal energy. The $H\alpha$ tracers are a subset of the total amount of bubble tracers, so they should thus contribute much less thermal energy. Meanwhile, the kinetic energy scales are about the same between the figures. This might suggest that the $H\alpha$ tracers, being on the edge of the bubbles as they expand through the ISM, carry more of the kinetic energy with them than hotter tracers in the interior of the bubble.

\subsection{Spatial Distribution} \label{subsec:spatial distribution}
\begin{figure}[ht!]
	\includegraphics[width=\linewidth]{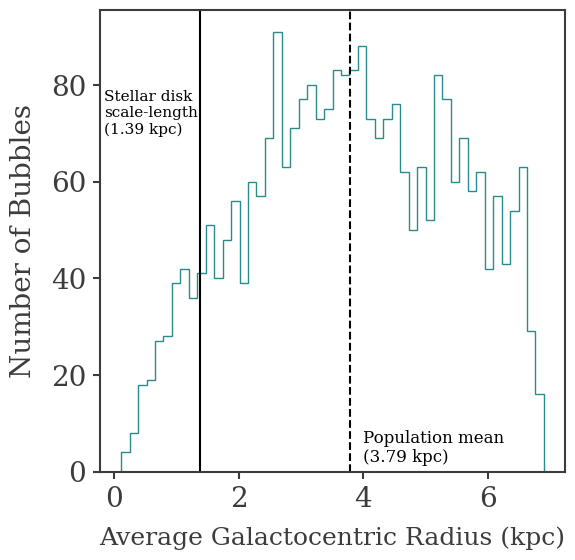}
	\caption{Bubble average galactocentric radius distribution. The median galactocentric radius is 3.79~kpc.}
	\label{Fig::galactocentric radius}
\end{figure}

We calculate the average galactocentric radii of the bubbles as the average distance from the bubble's centroid to the galactic center across its lifetime. We produce the distribution in Figure~\ref{Fig::galactocentric radius}, finding a median galactocentric radius of 3.79~kpc. We note that this median represents the delicate balance between the higher supernova activity toward the center of the galaxy and the larger galactic area for bubble formation at radial bins of higher distances.

\begin{figure}[ht!]
	\includegraphics[width=\linewidth]{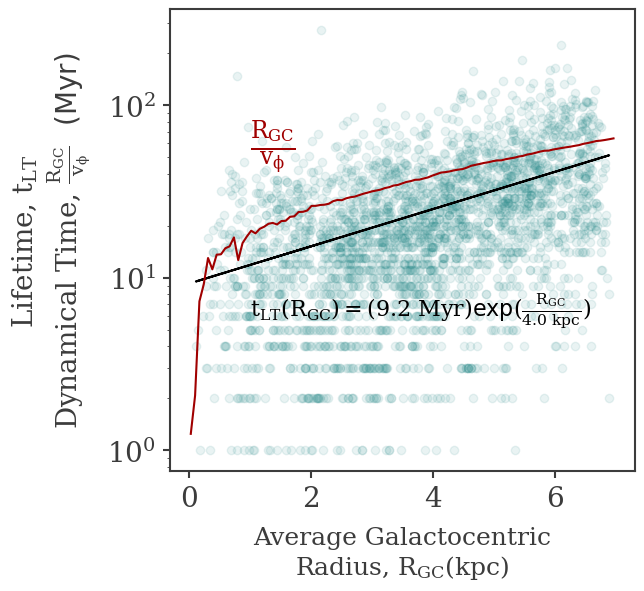}
	\caption{Bubble lifetime and dynamical time vs. average galactocentric radius. Bubbles tend to survive longer further from the galactic center. $\frac{R_{GC}}{v_{\phi}}$ is the median dynamical time of all the gas tracers in the simulation at one time step binned by galactocentric radius. The dashed purple line is an exponential trendline fit to the data. The uncertainties in the parameters of the fitted trendline are as follows: $9.2\pm0.5, 4.0\pm0.2$.}
	\label{Fig::lifetime vs galactocentric radius}
\end{figure}
We take the average galactocentric distance across each bubble's lifetime and plot the lifetime against it in Figure~\ref{Fig::lifetime vs galactocentric radius}. We find that there is a moderate positive exponential correlation between the lifetime and the galactocentric radius. The Spearman correlation coefficient $r_s$ is $0.46$. In an observational study, one possible explanation for this trend would be that the bubbles are subjected to fewer dynamic shearing forces in the outer edges of the galaxy, leading to longer bubble candidate connection chains and thus longer lifetimes. However, because bubbles identified with this algorithm die when they cool, the correlation in this simulated data is directly a result of temperature evolution, not shearing. Shearing could help the indentified bubbles cool faster by mixing the gas tracers into the surrounding ISM more rapidly.

\begin{figure}[ht!]
	\includegraphics[width=\linewidth]{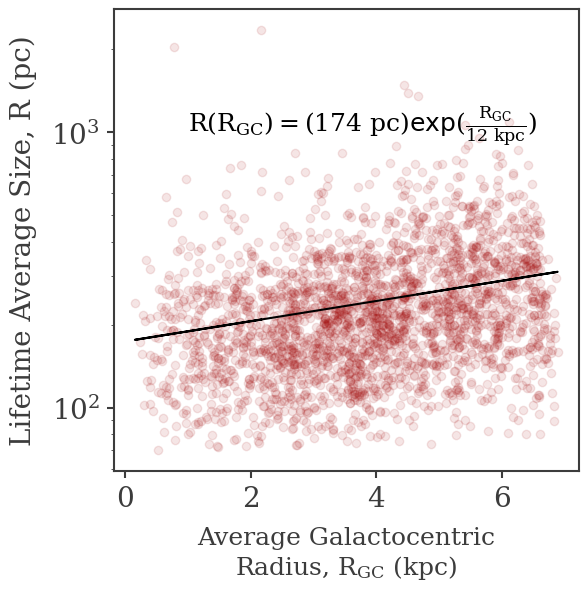}
	\caption{Lifetime average bubble size vs. average galactocentric radius. Larger bubbles are more commonly found further from the galactic center. The dashed purple line is an exponential trendline fit to the data. The uncertainties in the parameters of the fitted curve are as follows: $174\pm6, 12\pm1$.}
	\label{Fig::size vs galactocentric radius}
\end{figure}

This trend also supports the distribution in Figure~\ref{Fig::galactocentric radius}, particularly in higher radial bins. Longer bubble lifetimes imply longer chains of bubble candidate connections, leading to a lower new bubble formation rate as compared to regions of the galaxy with shorter bubble lifetimes. To test this, we take the averaged bubble sizes from Figure~\ref{Fig::size dist} and plot them against the average galactocentric radius in Figure~\ref{Fig::size vs galactocentric radius}. We find another slight positive exponential correlation between the size of the bubbles and the galactocentric radius as expected. In this case, $r_s = 0.27$. Although this positive correlation exists, upon sorting the bubbles by size, we find that the largest 50 bubbles do not seem to preferentially appear at higher galactocentric radii. In fact, some of these bubbles are closer to the galactic center than the population mean.
\begin{figure}[ht!]
	\includegraphics[width=\linewidth]{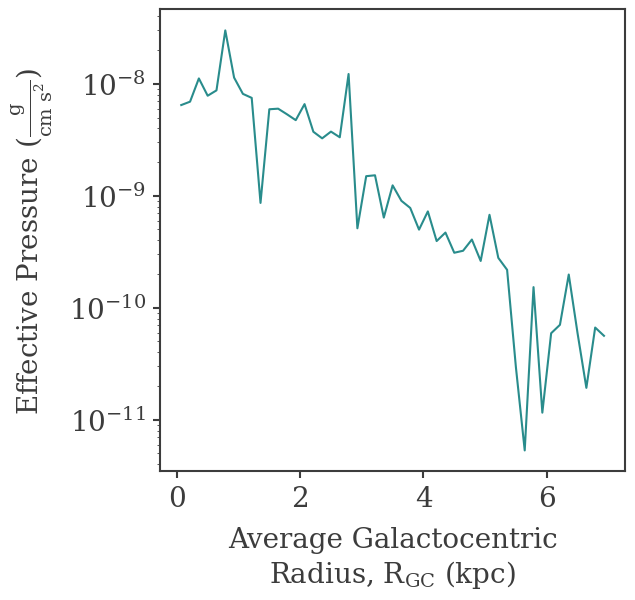}
	\caption{Effective pressure ($\rho\sigma^2$) vs. average galactocentric radius at $t=700$~Myr. Effective pressure is calculated in radial bins using all simulated gas tracers that aren't contained in identified bubbles.}
	\label{Fig::Effective Pressure vs Galactocentric Radius}
\end{figure}

Although the higher bubble lifetimes at farther galactocentric radii in Figure~\ref{Fig::lifetime vs galactocentric radius} may explain the larger sizes in Figure~\ref{Fig::size vs galactocentric radius}, physical conditions of the ISM should contribute as well. Larger bubbles may exist at higher galactocentric radii because of the lower external density of the surrounding ISM farther from the galactic center. At $t=700$~Myr, we calculate $\rho \sigma^2$ (the effective pressure) in radial bins, where $\rho$ is the density and $\sigma$ is the velocity dispersion of the gas tracers in a particular bin. We plot this property against galactocentric radius in Figure~\ref{Fig::Effective Pressure vs Galactocentric Radius} for all gas tracers not included in an identified bubble. This gas is a part of the external ISM surrounding the bubbles. We find that the effective pressure decreases at farther galactocentric radii, meaning that bubbles encounter less resistance to their expansion from the ISM in these less dense parts of the galaxy.

 We can additionally calculate the dynamical time for gas tracers at different galactocentric radii. We compute the tangential velocity of all gas tracers in the galaxy at an arbitrary timestep and bin them by their galactocentric radius. We take the median in each bin and then calculate $\frac{R_{GC}}{v_\phi}$ as a proxy for dynamical time. We plot this value in Figure~\ref{Fig::lifetime vs galactocentric radius}. This proxy for dynamical time increases at a remarkably similar rate to the lifetime with respect to galactocentric radii. We acknowledge that the lifetimes of the bubbles identified by this algorithm are directly linked to the cooling of their gas, not to shearing and destruction. However, shearing could help increase the rate of cooling for the bubbles by allowing hot gas to mix more rapidly with the surrounding ISM and cool; the increase in dynamical time at larger galactocentric radii could subsequently suggest slower mixing (and cooling) of hot gas tracers. Given that our proxy for dynamical time is about $\frac{1}{2\pi}$ times the expected orbital period, and that the lifetime is about a factor of 2 smaller than the proxy, we can predict that the average bubble completes on the order of $\frac{1}{10}$ of an orbit around the galaxy. The bubbles tend to complete a slightly larger fraction of their orbit at higher galactocentric radii.

\section{Discussion}
\label{sec:Discussion}
We have shown in Section~\ref{sec:bubble properties} that there exists a population of 2778 large, hot, ionized, feedback-driven bubbles in our dwarf spiral galaxy simulation. These bubbles expand and cool over time and can live up to $150$~Myr, with an average lifetime of $25$~Myr.

We note that the above results depend on the resolution of our simulation, as well as a number of other assumptions made in our modeling of star formation, feedback, and interstellar medium physics. In this section, we compare our results to studies of superbubble evolution in the literature and outline the key caveats of our model, along with their possible effects. We also give an overview of our planned future work using these results.

\subsection{Caveats of Our Model} \label{subsec:caveats}

\subsubsection{Size Resolution} \label{subsubsec:size resolution}
The mass resolution of the simulation imposes a size resolution that limits the range of bubbles that we can identify. The following equation approximates our radial resolution limit of the bubbles:
\begin{equation} \label{Eqn::Resolution}
R = (\frac{3MN_{tracers}}{4\pi\rho})^{\frac{1}{3}}
\end{equation}
where $M$ is the mass resolution of the simulation ($450~\rm{M}_{\odot}$), $N_{tracer}$ is the chosen tracer resolution limit (10), and $\rho$ is the typical density of the interior of a bubble, taken as $3.16 \times 10^6~\frac{\rm M_\odot}{kpc^3}$ after converting from particle number density (see Section~\ref{subsec:temperature and density}). We find a radial resolution limit of 70~pc, which is denoted in Figure~\ref{Fig::size} by the bottom dashed line on each plot.

Because superbubbles are typically defined on the scale of 100s of pc \citep{BubbleInfo}, the size limit of 70~pc essentially requires that all of our identified bubbles are superbubbles. This is consistent with our simulation's stellar feedback mechanisms. Supernova feedback is the dominant feedback mechanism in the simulation, and it is largely responsible for the production of superbubbles in observations.

We identify smaller bubble-like structures before filtering by tracer count, but we choose to discard these structures. Their small tracer counts make property analysis too volatile. Measuring the size of a bubble using only one tracer is impossible.

\subsubsection{Missing ISM Physics}

Our simulation did not include magnetic fields or cosmic rays. As shown in the TIGRESS simulations, these phenomena play an important role in shaping the ISM \citep{TIGRESS}. Their exclusion from our simulation may have important effects on bubble evolution.

Magnetic fields have a significant effect on the dynamics of expanding superbubbles, capable of decelerating the superbubble shock front \citep{MagneticFieldsShells}. They may also affect how the front fragments as it expands. Generally, magnetic fields have been shown to suppress bubble expansion and volume, even confining them to the galactic plane \citep{TomisakaMagBubbles}. As such, the bubble population we find might be slightly larger in radius because magnetic fields are not helping to slow their expansion.

Conversely, cosmic rays are partially responsible for the expansion of bubbles, so their exclusion from our simulation may decrease the size of some of the identified bubbles. Cosmic rays help supernova feedback create stronger galactic outflows \citep{PeschenCROutflows}, which would manifest themselves as larger bubbles if included in our simulation. Without their momentum injection, the driving expansion force for the bubbles would rest solely on momentum injection from supernovae and stellar winds.

\subsection{Comparison to the Literature}

A recent study follows a procedure similar to ours to identify stellar-feedback driven superbubbles in an {\sc Arepo} galaxy simulation and analyze their time-evolving properties \citep{LiSimBubbles}. However, before comparing our results, there are key distinctions between our methodologies that must be discussed. Li et al. utilize a different bubble identification algorithm, focusing on the simulation cells rather than passive gas tracers as in our algorithm. This allows them to achieve a spatial resolution of 4~pc. As a result, they are able to explore the properties of much smaller bubbles, whereas we are limited to superbubbles larger than 70~pc in radius.

With these caveats in mind, \cite{LiSimBubbles} produces a remarkably similar superbubble size distribution to our results. Our lifetime average bubble size distribution peaks around 200~pc, as does theirs. Both distributions also decrease rapidly as the bubble size approaches 1000~pc. Interestingly, when Li et al. add in the other sources of stellar feedback on top of supernova feedback, they find a secondary peak in their size distribution for bubbles smaller than 100~pc. If we were able to resolve smaller bubbles, we may also expect to see a similar secondary peak for the smaller bubbles.

Our bubble size evolutions are slightly more difficult to compare. \cite{LiSimBubbles} finds that the majority of their bubbles expand into 100s of pc in size over their first $10$~Myr of life. As seen in Figure \ref{Fig::size}, we generally find that our bubbles evolve similarly. Any discrepancy between our results might arise from the bubbles we considered; Li et al. only focus on the main progenitors and descendants in their bubble network, or the candidates that share the most mass. Meanwhile, we consider all candidates that have interacted with a bubble. This likely appears as a rapid expansion in bubble size as we conglomerate multiple hot candidates to a bubble during the beginning of a bubble's lifetime while Li et. al. only track one candidate per time step.

We also compare our bubble energies to that expected by other work. The energy released by a supernova is about $10^{51}$~erg. About 1-10\% of this energy is retained by the ISM as kinetic and thermal energy \citep{Walch2015} The spread of kinetic energies in Figure~\ref{Fig::energy} reflect these expectations nicely--most bubbles have less than $10^{50}$~erg of energy. The bubbles on the higher end of energies are the most massive superbubbles that have spread across the galaxy and accumulated excess gas tracers. These bubbles are also more likely to have experienced many supernovae due to their larger size. Similarly, those at the lowest energies are likely old bubbles with few tracers left that are cooling and mixing back into the ISM.

We also present statistical observational studies to compare with our results. In a recent PHANGS survey, \cite{WatkinsEnergy} observe a mean bubble radius of 134~pc in nearby galaxies. In Figure \ref{Fig::size dist}, we find a mean bubble radius of 251~pc. However, in this survey, Watkins et al. resolve bubbles down to 30~pc, and their largest detected bubble has a radius of 330~pc. Meanwhile, our size resolution limit is 70~pc, and our largest lifetime averaged bubble size is over 2000~pc in radius. We find that our size distribution is skewed towards much larger bubbles. The differences in our identification methods might cause this; an older, larger bubble spread across kpc scales may be impossible to identify by eye. Meanwhile, in our identification method, since the gas tracers of a bubble remain linked as long as they are above the temperature threshold, we do not need to question whether distant remnants of a single bubble are connected, and we continue to treat them like a connected bubble. In another PHANGS study of NGC628, \cite{WatkinsMain} find bubbles ranging in size from 6-552~pc. Our largest measured bubbles far surpassed 552~pc for the aforementioned reasons. They also find a similar slight positive correlation between bubble size and galactocentric radius as in Figure \ref{Fig::size vs galactocentric radius}.

Several statistical observational studies have shown the potential impact of superbubbles on star formation. For instance, stellar feedback has been shown to have a positive effect on molecular gas fraction in superbubbles in the LMC \citep{DawsonLMC} and the Milky Way \citep{DawsonMilkyWay}. Colliding superbubbles can also create dense, molecular filaments \citep{TannerMolecularFilaments}. Increasing the molecular gas fraction of the ISM around superbubbles primes these regions for new star formation, so superbubbles can have a significant effect on galactic star formation.

We can even see these effects in our Local Neighborhood. The surface of our own Local Bubble contains almost all of the closest star forming regions which are expanding radially outward \citep{ZuckerLocal}; this is evidence of supernova feedback influencing these molecular clouds. Similarly, the Perseus and Taurus molecular clouds lie along the spherical Per-Tau shell, another supernova-driven bubble \citep{Bialy2021}. Barnard's Loop was likely shaped by stellar feedback, and all three of Orion's molecular clouds fall along the Orion shell \citep{Foley2023}. Gaia data correspondingly shows that many molecular clouds are found along the edges of superbubbles \citep{ZuckerReview}. All of these nearby star-forming regions being found around bubbles and regions of stellar feedback is great evidence that stellar feedback is partially responsible for triggering new star formation.

All of these examples of linked stellar feedback and star formation both locally and in extragalactic environments demonstrate the importance of developing bubble identification methods. With the tools to identify and analyze the evolution of bubbles and their surrounding environment, we can explain how they subsequently trigger star formation on galactic scales. In future work, we plan to link our own bubble catalog to star particles at their expanding shock fronts to  determine if the bubbles trigger further star formation.

\section{Conclusions}
\label{sec:Conclusions}
Supernova-driven bubbles profoundly affect the structure of gas in the galactic disk. Their widespread galactic presence becomes more evident as we receive new data from JWST and information about our own Local Bubble. Identification of bubbles in galaxy simulations thus presents a special opportunity to learn about the dynamics of bubble formation and evolution. We present our key findings below.
\begin{enumerate}
    \item We propose a novel, successful physical bubble-identification and tracking algorithm, generating a list of 2778 time-evolving bubbles for analysis over 300 Myr of galactic evolution.
    \item We find an exponential bubble lifetime distribution with a half-life of 18~Myr and mean lifetime of 25~Myr. These lifetimes are in good agreement with observational estimates of superbubble lifetimes in the local Milky Way.
    \item We find an exponential distribution of bubble sizes that typically grow in size over the lifetime of the bubbles. These sizes closely match observed bubble radii in other galaxies from JWST.
    \item We find a median average galactocentric bubble radius of 3.79~kpc with a trend toward higher lifetimes and larger sizes at higher galactocentric radii. This suggests that bubbles can grow larger and survive longer in the outer reaches of a galaxy.
    \item We propose the use of warm gas tracers as a synthetic observation of H$\alpha$. Using H$\alpha$ as a tracer still misses a large fraction of superbubble evolution at early times.
\end{enumerate}

The vast majority of the bubbles have lifetimes less than 150~Myr. Because the bubble lifetimes follow an exponential distribution, it is most common for bubbles to have very short lifetimes. The bubbles that survive the longest often accomplish this because they are extremely large and can live on by sharing their mass with many bubble candidates.

Similarly, very few bubbles are larger than 1 kpc in radius. The bubble sizes are described well by an exponential distribution, with most bubbles being smaller. The largest bubbles also tend to survive the longest through the same mass-sharing described above.

We demonstrate the usefulness of our identification techniques by determining a range of bubble properties. Beyond theory, our work offers valuable guidance for the interpretation of observational data and illuminates the broad trends in bubble evolution. Moreover, our data provide a much fuller understanding of the lifetimes and the potential subsequent lasting impacts of the bubbles.

\makeatletter
\let\linenumbers\relax
\let\internallinenumbers\relax
\makeatother

\begin{acknowledgments}
\section*{Acknowledgments}
We thank Eric Koch, Theo O'Neill, and Gus Beane for their advice and guidance in bringing together this work. A.A. would like to thank the Northeastern College of Science for supporting this work through the COS Unpaid Domestic Research Co-op Scholarship. S.M.R.J. and M.M.F. wish to express their gratitude to the Smithsonian Scholarly Studies Fund for supporting this work.  S.M.R.J. is supported by Harvard University through an Institute of Theory and Computation Fellowship.

Our visualization and exploration of data would not have been possible without \texttt{glue} visualization software. \textit{Software:} \texttt{glue} \citep{Robitaille2017-af}, \texttt{numpy} \citep{Harris2020-cd}, \texttt{scipy} \citep{Virtanen2020-jp}, \texttt{COLT} \citep{COLT}, \texttt{AREPO} \citep{Springel10}.
\end{acknowledgments}

\bibliography{main}{}
\bibliographystyle{aasjournal}

\end{document}